\documentstyle[sprocl,epsf,overcite,refsty_p]{article}

\newcommand{\id}{\mbox{d}}

\newcommand{\GeVs}{\mbox{GeV}^2}
\newcommand{\beq}{\begin{equation}}
\newcommand{\eeq}{\end{equation}}
\begin{document}

\titlepage
 
\vbox to 0pt{
\vspace*{-2.5cm}
\begin{flushright}
\sf
{\large SMC/97/06}\\
 
\medskip
hep-ex/9704005\\
March, 1997
\end{flushright}
\vss}

\title{SPIN STRUCTURE FUNCTIONS}

\author{ G.K. MALLOT }

\address{
Institut f\"ur Kernphysik der Universit\"at Mainz, Becherweg 45,\\
D-55099 Mainz, Germany}


\maketitle\abstracts{
The experimental status of the spin-dependent structure functions as 
obtained from the deep inelastic scattering experiments at CERN, SLAC, 
and DESY is reviewed.
All data show a violation of the Ellis--Jaffe sum rule.
The Bjorken sum rule is found to be valid and is tested to the 10~\% level.
}

\vfill

\begin{center}
Invited plenary talks given at the \\
\medskip
\it
XVI International Conference on Physics in Collision, PIC96\\
\rm
Mexico City, Mexico, 19--21 June, 1996\\
\medskip
and\\
\medskip
\it
12th International Symposium on High-Energy Spin Physics, SPIN96\\
\rm
Amsterdam, The Netherlands, 10--14 September 1996\\

\bigskip
\it
to be published in the proceedings \footnote{
To the SPIN96 proceedings a shorter version of this paper was submitted.} 
\end{center}
\vfill
 
\newpage
\titlepage
~~~~
\newpage
\setcounter{page}{1}
 
\title{SPIN STRUCTURE FUNCTIONS}
 
\author{ G.K. MALLOT }
 
\address{
Institut f\"ur Kernphysik der Universit\"at Mainz, Becherweg 45,\\
D-55099 Mainz, Germany\\
email: Gerhard.Mallot@cern.ch}
 
 
\maketitle\abstracts{
The experimental status of the spin-dependent structure functions as
obtained from the deep inelastic scattering experiments at CERN, SLAC,
and DESY is reviewed.
All data show a violation of the Ellis--Jaffe sum rule.
The Bjorken sum rule is found to be valid and is tested to the 10~\% level.
}

\section{Introduction}

After the discovery by the EMC~\cite{EMC88a,EMC89a} that the contribution of 
the quark spins to the proton spin is much smaller than expected,
new experiments were proposed and set up to study this ``spin puzzle''.
First was the experiment of the Spin Muon Collaboration (SMC) at 
CERN~\cite{SMC94b,SMC94c,SMC95a,SMC96a,SMC97a,SMC96z} followed by
the SLAC experiments~\cite{E142_96,E143_95a,E143_95b,E143_96a,E154_96a} 
and the HERMES experiment~\cite{HER97a} at DESY.
In parallel intensive work started on the theoretical side.
The information obtained from the measurements is twofold. 
On one hand the test of the Bjorken sum rule~\cite{Bjo66} provides a sensitive
test of QCD, on the other hand the nonperturbative spin-flavour structure of the
nucleon is contained in the polarised parton distribution functions.
The structure function data are by now precise enough to perform sensible 
QCD analyses and to determine the parton distribution functions.
However, the precision of the data is still far from the quality of the 
spin-averaged structure function measurements and several constraints are needed
in the QCD analyses.
An important aim is to understand the nucleon's spin in terms of the spins
of quarks and gluons, 
$\Delta\Sigma=\Delta u + \Delta d + \Delta s$ and $\Delta g$, and their 
orbital angular momenta, $L_q$ and $L_g$,
\beq
\label{eq:heliSR}
\frac{1}{2}=\frac{1}{2}\Delta\Sigma+L_q+\Delta g + L_g.
\eeq
It is well known from both, experiment and theory, that at high $Q^2$ about 
half of the nucleon's longitudinal momentum is carried by the gluons. 
Recently, it was predicted\cite{JiT96a} that the same sharing should 
apply for the total angular momentum.
In the Quark Parton Model the quark polarisations,
\beq
\Delta q = \left(q^{+}-q^{-}\right) + 
           \left(\overline{q}^{+}-\overline{q}^{-}\right),
\eeq 
are related to the spin-dependent structure function $g_1$ by
\beq
g_1(x,Q^2) = \frac{1}{2}\sum_{f} e_f^2 \Delta q_f(x,Q^2),
\eeq
where $f$ runs over the quark flavours and $e_f$ are the electrical
quark charges. The notations $q^{+(-)}$ refer to
parallel (antiparallel) orientation of the quark and nucleon spins.

Experimentally the spin-dependent structure functions, $g_1$ and $g_2$, 
are obtained from the measured event-number asymmetries,
$A^{\rm raw}_\parallel$, for longitudinal orientation of the target and
lepton spins, and $A^{\rm raw}_\perp$ for transverse target polarisations.
These raw asymmetries range for the proton typically from
a few per cent at large $x$ to a few parts per thousand at small $x$.
In the lepton-nucleon asymmetries, $A_\parallel$ and $A_\perp$,
the uncertainties of the raw asymmetries get amplified by the factor 
$1/P_bP_tf$ accounting for the incomplete beam and target polarisations, 
$P_b$ and $P_t$, and the dilution factor, $f$.
Typical target materials contain a large fraction of unpolarisable nucleons
and $f$ denotes the fraction of the total spin-averaged cross section arising 
from the polarisable nucleons.
For the target materials used, $f$ varies from about 0.13 (butanol),
over 0.17 (ammonia) to 0.3 ($^3$He). For deuterated butanol and ammonia
$f$ is 0.23 and 0.3, respectively, while for the proton and deuteron gas 
targets of the HERMES experiment $f$ is close to unity.
The neutron structure functions are either obtained from the combination of
proton and deuteron data or from experiments using $^3$He targets. 
The deuteron asymmetries are
slightly reduced from the average of proton and neutron asymmetries due to the
D-state component in the deuteron wave function.
The $^3$He asymmetry is mainly due to the unpaired neutron, 
however a small proton contribution has to be corrected for.
Due to the cancellation of the isotriplet part in $g_1^{\rm d}$
the measurements using deuteron targets are most sensitive to the
flavour-singlet part, $\Delta\Sigma$, and thus to the violation of the
Ellis--Jaffe sum rule.
The structure functions, $g_1$ and $g_2$, are related to the virtual photon
asymmetries,  $A_1$ and $A_2$, via
\begin{eqnarray}
\label{eq:ApAt}
A_\parallel=&D(A_1+\eta A_2),  \hskip 1cm & A_\perp=d(A_2-\xi A_1),\\
\label{eq:A1A2}
A_1=&\displaystyle\frac{g_1-\gamma^2 g_2}{F_1},  \hskip  1cm & A_2=\gamma\frac{g_1+g_2}{F_1}.
\end{eqnarray}
Here $F_1=F_2(1+\gamma^2)/2x(1+R)$
is the well-known spin-averaged structure function and
$\gamma^2=Q^2/\nu^2$, $\eta$, and $\xi$ are kinematic factors, which
are small in most of the kinematic domain covered by the data.
The variables $\nu$ and $Q^2=-q^2$ denote respectively the energy transfer
and negative square of the 4-momentum transfer.
The kinematic factors, $D$ and $d$, account for the incomplete
transverse polarisation of the virtual photon. 
With longitudinal target polarisation predominantly $g_1$ is determined, while
experiments with transverse target polarisation are sensitive to $g_1+g_2$.
The virtual photon asymmetries are bounded by $|A_1|\le 1$ and $|A_2|\le\sqrt{R}$,
where $R=\sigma_L/\sigma_T$ is the longitudinal-to-transverse photoabsorption
cross-section ratio known, like $F_2$, from unpolarised deep inelastic scattering.

\section{The Experiments}

\begin{figure}[t]
\begin{center}
      \mbox{\epsfxsize=0.65\hsize\epsfbox[0 0 604 473]{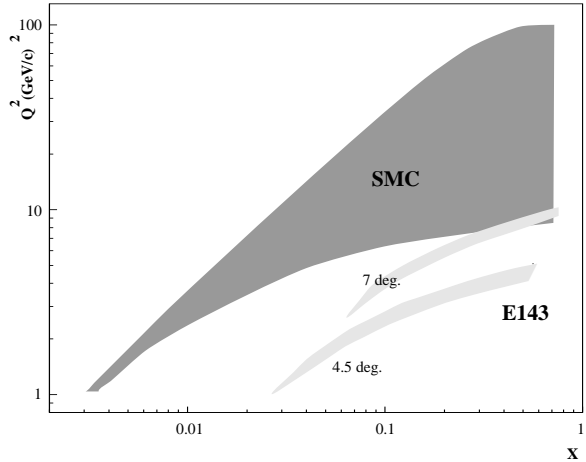}}
      \caption{
            The kinematic ranges covered by the SMC and E-143 experiments.}
      \label{fig:range}
      \end{center}
      \end{figure}

New data from the 1995 runs come from the SMC\cite{SMC97a} (deuteron), 
the HERMES experiment\cite{HER97a} ($^3$He) and the SLAC experiment 
E-154\cite{E154_96a} ($^3$He, preliminary).
Some important parameters of these and the earlier experiments 
are summarised in Table~\ref{tab:exp} and the kinematic ranges of the SMC 
and E-143 experiments are compared in Fig.~\ref{fig:range}.

\begin{table}
\begin{center}
\caption{Parameters of the experiments}
\label{tab:exp}
\begin{tabular}{llrccl@{$\,\le\,$}c@{$\,\le\,$}lc}
\hline
\hline
Experiment  &Lab.  &\multicolumn{2}{c}{Beam}& $\langle Q^2\rangle$&
 \multicolumn{3}{c}{$x$ range}& Targets\\
            &      &&                       &     [GeV$^2$]         & 
\multicolumn{3}{c}{} &\\
\hline
E-80/130 &S\sc lac &  22~GeV&e    & ~~4~~ & 0.18  &$ x $& 0.7& p \\
EMC      &C\sc ern & 200~GeV&$\mu$&  10.7 & 0.01  &$ x $& 0.7& p \\
SMC      &C\sc ern & 190~GeV&$\mu$&  10~~ & 0.003 &$ x $& 0.7& p, d \\
E-142/143&S\sc lac &  29~GeV&e    & ~2--3~& 0.03  &$ x $& 0.8& $^3$He, p, d \\
E-154/155&S\sc lac &  48~GeV&e    & ~~5~~ & 0.014 &$ x $& 0.7& $^3$He, p, d \\
HERMES   &D\sc esy &  27~GeV&e    & ~~3~~ & 0.023 &$ x $& 0.6& $^3$He, p, d \\
\hline
\hline
\end{tabular}
\end{center}
\end{table}

\bigskip
The experiments of the SMC were performed at the CERN muon beam
at 190~GeV using a large, double-cell solid-state polarised target. 
Data with parallel and antiparallel orientation of the lepton and target spins were 
taken simultaneously. 
This technique cancels most systematic uncertainties and overcomes the problem that 
the natural polarisation of the muon beam cannot effectively be reversed.
Average polarisations of 86~\% and 50~\% were achieved for the butanol and deuterated 
butanol targets used in the years 1992--1995.
The opposite polarisation of the material in the two 65~cm long cells was reversed 
every 5~h. 
In the 1996 run a proton target made of ammonia was employed.
The incoming and scattered muons were analysed by magnetic spectrometers. 
Hadrons produced in the scattering were also detected.
However, a particle identification beyond electron-hadron separation was not 
attempted.
A dedicated spectrometer downstream of the main scattering spectrometer measures the
about 80~\% polarisation of the muon beam by two methods.
One is based on the dependence
of the energy spectrum of the decay positrons on the parent muon polarisation, 
the other uses the asymmetry in polarised muon-electron scattering from
a magnetised iron foil.
The domain unique to the SMC experiment is high $Q^2$ and small $x$ which is 
essential
for the extrapolations to $x\rightarrow0$ and for QCD analyses of the structure
function data.

The SLAC experiments E-142/E-143 and E-154 were performed at the SLAC electron beam
in End Station~A with energies up to 29 and 48~GeV, respectively. 
The rather low energy of the incoming electron limits the accessible kinematic range. 
Apart from the E-142 experiment (39~\%) electron polarisations of 80~\% were reached 
using strained-lattice GaAs photocathodes. 
The electron polarisation was varied randomly on a pulse-by-pulse basis
and measured by Moller scattering from thin ferromagnetic foils.
For the neutron experiments, E-142 and E-154, a $^3$He gas target with a 
pressure of 10~bar was used and
polarisations of 30--40~\% were reached by spin-exchange with optically pumped rubidium vapour.
With the ammonia targets used in the E-143 experiment and foreseen for the E-155 experiment
polarisations of 65--80~\% for NH$_3$ and 25~\% for ND$_3$ were reached.
For the E-155 experiment running in 1997 also a $^6$LiD deuteron target is foreseen.
The lithium-6 nucleus can be understood as an $\alpha$+d system~\cite{Sch93}, 
yielding for $^6$LiD a favourable dilution factor of $f=0.5$.
The scattered electrons were analysed by two magnetic spectrometers placed under
different scattering angles to increase the acceptance. 
With the intense electron beams much higher luminosities than with muon beams can be
achieved yielding smaller statistical errors. 
The accuracy of the first moments obtained from the SLAC experiments is thus 
systematics limited.

A different technique is applied in the HERMES experiment which started 
data-taking in 1995.
The electrons in the HERA ring self-polarise by the Sokolov--Ternov effect to about
50~\%. Spin rotators before and after the target provide longitudinal
electron polarisation.
A 40~cm long windowless storage cell placed in the ring is filled with up 
to 10$^{17}$ atoms/s which are then pumped away at both ends of the cell. 
Thus pure polarised proton, deuteron, and $^3$He targets can be used without diluting
the asymmetries by the presence of unpolarised target materials. 
This target technique is in particular ideal to study hadrons produced in the deep 
inelastic scattering process.
The polarisation of the hydrogen and the deuterium target can be changed within 
milliseconds while obtaining equilibrium in the $^3$He target takes about 20~s. 
With the $^3$He target employed in the 1995 run polarisations of 50~\% were reached.
The scattered electrons are analysed in an open magnetic spectrometer.
Particle identification is provided by the combined information from a TRD, a 
threshold Cherenkov counter, and a lead-glass calorimeter.
Due to the limited energy of the HERA electron ring of 27~GeV HERMES cannot extend the
kinematic range of the SMC and SLAC experiments. 
However, it will provide high-precision semi-inclusive data.

\section{Asymmetries and Structure Functions}
\begin{figure}[t]
\begin{center}
      \mbox{\epsfxsize=0.65\hsize\epsfbox[12  7 538 828]{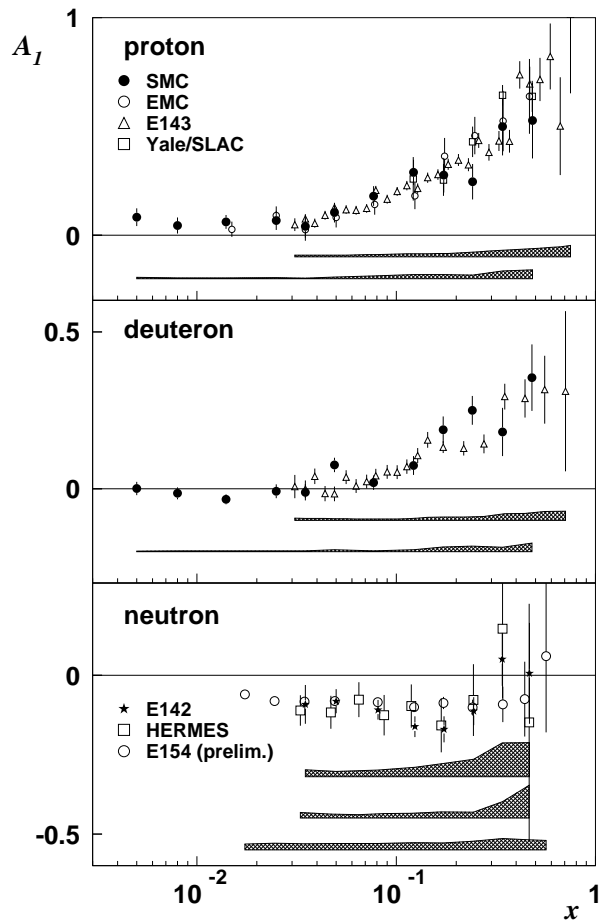}}
      \caption{
            The virtual-photon asymmetry, $A_1$, as a function of $x$.
            The shaded bands indicate the systematic errors,
            from top to bottom: proton: E-143, SMC; deuteron: E-143, SMC;
            neutron: E-142, Hermes, E-154.}
      \label{fig:a1all}
   \end{center}
   \end{figure}

The virtual-photon asymmetries measured by the various experiments with proton,
deuteron, and neutron targets are summarised for $Q^2>1~\GeVs$ in Fig.~\ref{fig:a1all}.
Although the SMC data were taken at an about six-times higher value of $Q^2$ than
the SLAC E-143 data the agreement between the two data sets is excellent.
Thus, in the region of overlap, the data are compatible with no $Q^2$ dependence of 
$A_1$. 
Asymmetries for $Q^2<1~\GeVs$ were published by the E-143 Collaboration\cite{E143_95c} 
and the SMC\cite{SMC96z,SMC97a}. However, due to higher-twist contributions they are
difficult to interpret.
\begin{figure}[t]
   \begin{center}
      \mbox{\epsfxsize=0.65\hsize\epsfbox[12  7 538 828]{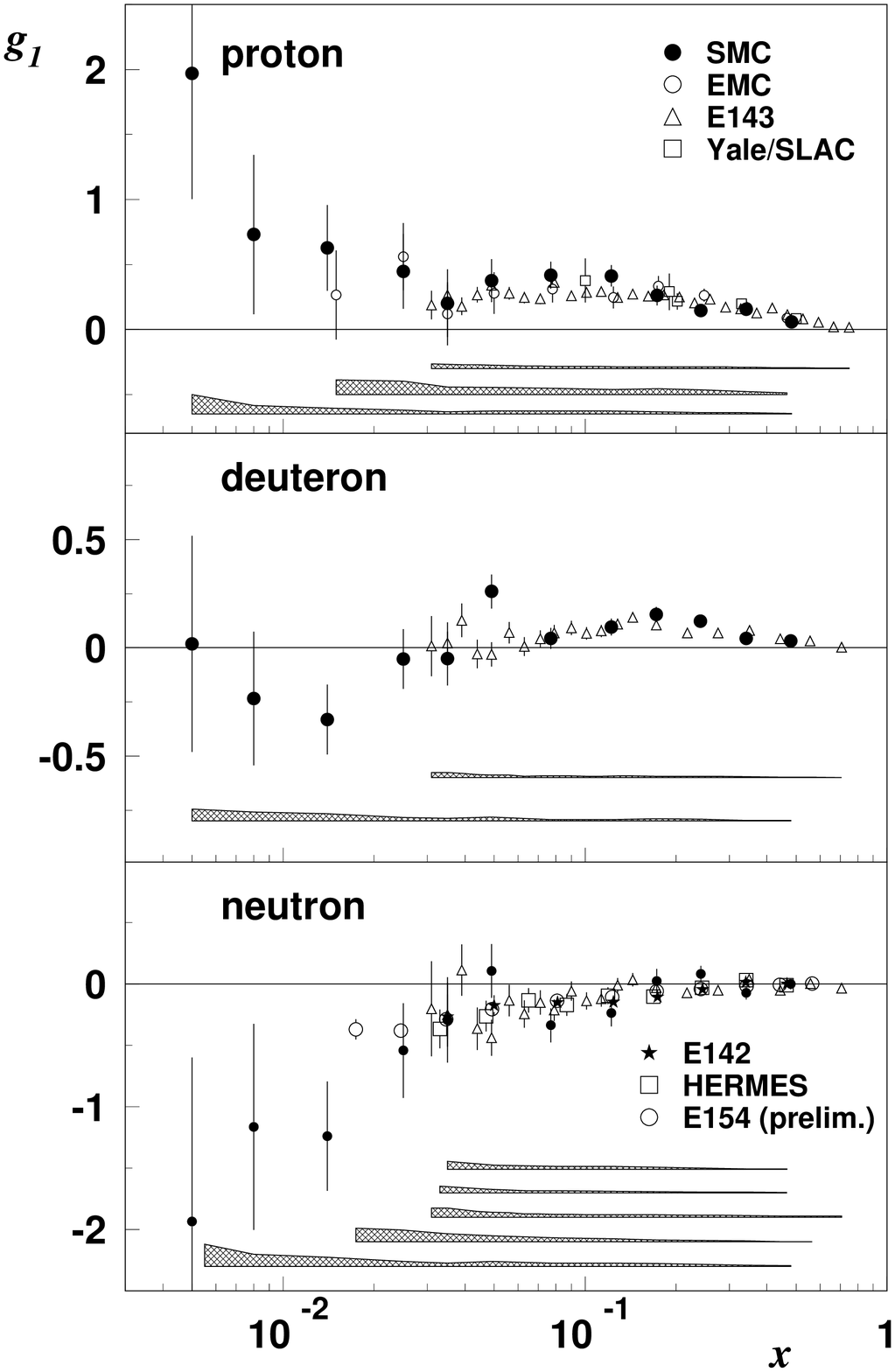}}
           \caption{
            The structure function $g_1(x,Q^2)$ as a function of $x$ at the 
            $Q^2$ of the individual measurements. The systematic errors are 
            indicated by the shaded bands,
            from top to bottom: proton: E-143, EMC, SMC; deuteron: E-143, SMC;
            neutron: E-142, Hermes, E-143, E-154, SMC. The E-143 and SMC neutron
            data are calculated from the proton and deuteron data.
            }
      \label{fig:g1all}
   \end{center}
   \end{figure}
From the asymmetries and Eqs.~\ref{eq:A1A2} the structure function $g_1$ is obtained as 
\beq
\label{eq:g1}
g_1(x,Q^2) = \frac{F_2(x,Q^2)}{2x\left(1+R(x,Q^2)\right)}
             \left(A_1(x,Q^2)+\gamma A_2(x,Q^2)\right).
\eeq
Usually the term $\gamma A_2$ is neglected, what is consistent 
with experimental results for $A_2$~\cite{SMC96z,SMC97a,E143_96a}.
In the analysis of some SLAC experiments, where $\gamma^2$ is larger than in
the high-energy muon experiments, Eqs.~\ref{eq:ApAt} were solved for both, $A_1$ and $A_2$.
In all analyses the parametrisations of $F_2$ by the NMC~\cite{NMC92e,NMC95c} 
and of $R$ by SLAC~\cite{WhR90} were used. New $R$ data from the NMC~\cite{NMC96a}
for $x<0.1$ agree well with $R_{\rm QCD}$, where $R_{\rm QCD}$ is 
calculated~\cite{AlM78} in perturbative QCD using the experimental gluon 
distribution function. In this $x$ region, previously basically uncovered by 
experiments, the SLAC $R$ parametrisation deviates somewhat from the data.
Note that apart from radiative corrections and $Q^2$ extrapolations
the $R$ dependence cancels in the experimental $g_1$, when the same values 
of $R$ were used in the $g_1$ and $F_2$ analyses. For a recent review of the
unpolarised structure functions with emphasis on the fixed-target experiments
see Ref.~[\citen{Mal97a}].

\begin{figure}[t]
\begin{center}
\mbox{\epsfxsize=0.65\hsize\epsfbox[0 0 390 616]{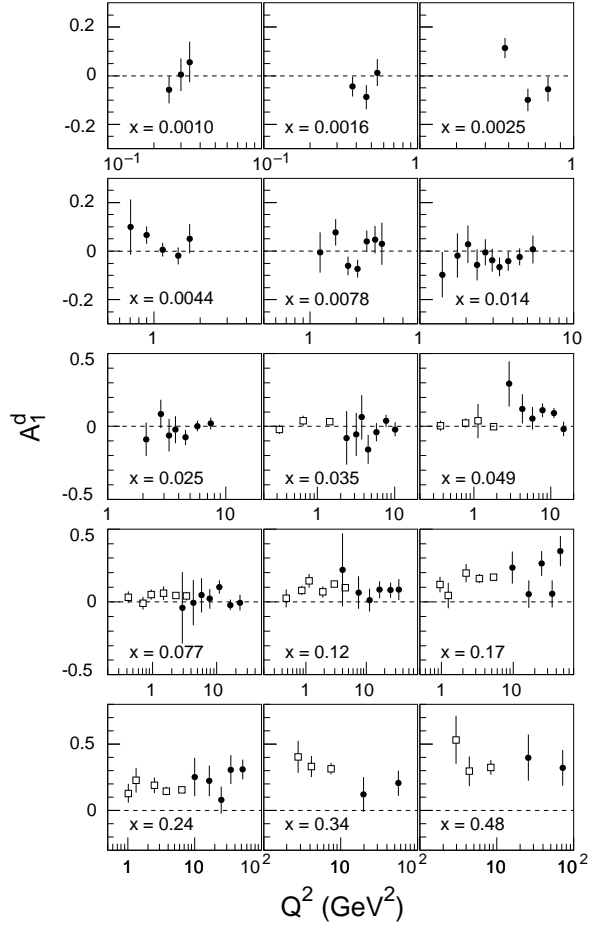}}
\end{center}
\caption{
Virtual-photon asymmetry of the deuteron, $A_1^{\rm d}$, as a function of $Q^2$ in several $x$ bins
from the SMC\protect\cite{SMC97a} (full circles) and E-143\protect\cite{E143_95c}
(open squares).}
\label{fig:a1dxq}
\end{figure}

The structure function data at the
$Q^2$ of the individual measurements are shown in Fig.~\ref{fig:g1all}.
The values of $Q^2$ vary from about 1~GeV$^2$ for the smallest-$x$ point of each 
data set to typically 
50~GeV$^2$ (10~GeV$^2$) for the largest-$x$ point of the SMC (SLAC) data.  
A study of the $Q^2$ dependence of $A_1$ using the combined SMC/EMC and SLAC
data was first published~\cite{E143_95c} by the E-143 Collaboration.
No significant $Q^2$ dependence was found for $Q^2>1~\GeVs$. 
For this analysis only one SMC data point per $x$-bin at an average $Q^2$ was 
available.
A compilation of all now available deuteron asymmetries (SMC/E-143) is shown 
in Fig.~\ref{fig:a1dxq}. 
The proton data are in precision and kinematic coverage similar to the deuteron 
data. Still
the experimental precision is not sufficient to reveal any $Q^2$ 
dependence of $A_1$.
It has to be noted that in the small-$x$ region, where rather large effects are
expected, the lever arm in $Q^2$ is very limited.

\begin{figure}
\begin{minipage}[t]{0.47\hsize}
\mbox{\epsfxsize=\hsize\epsfbox[20 10 538 538]{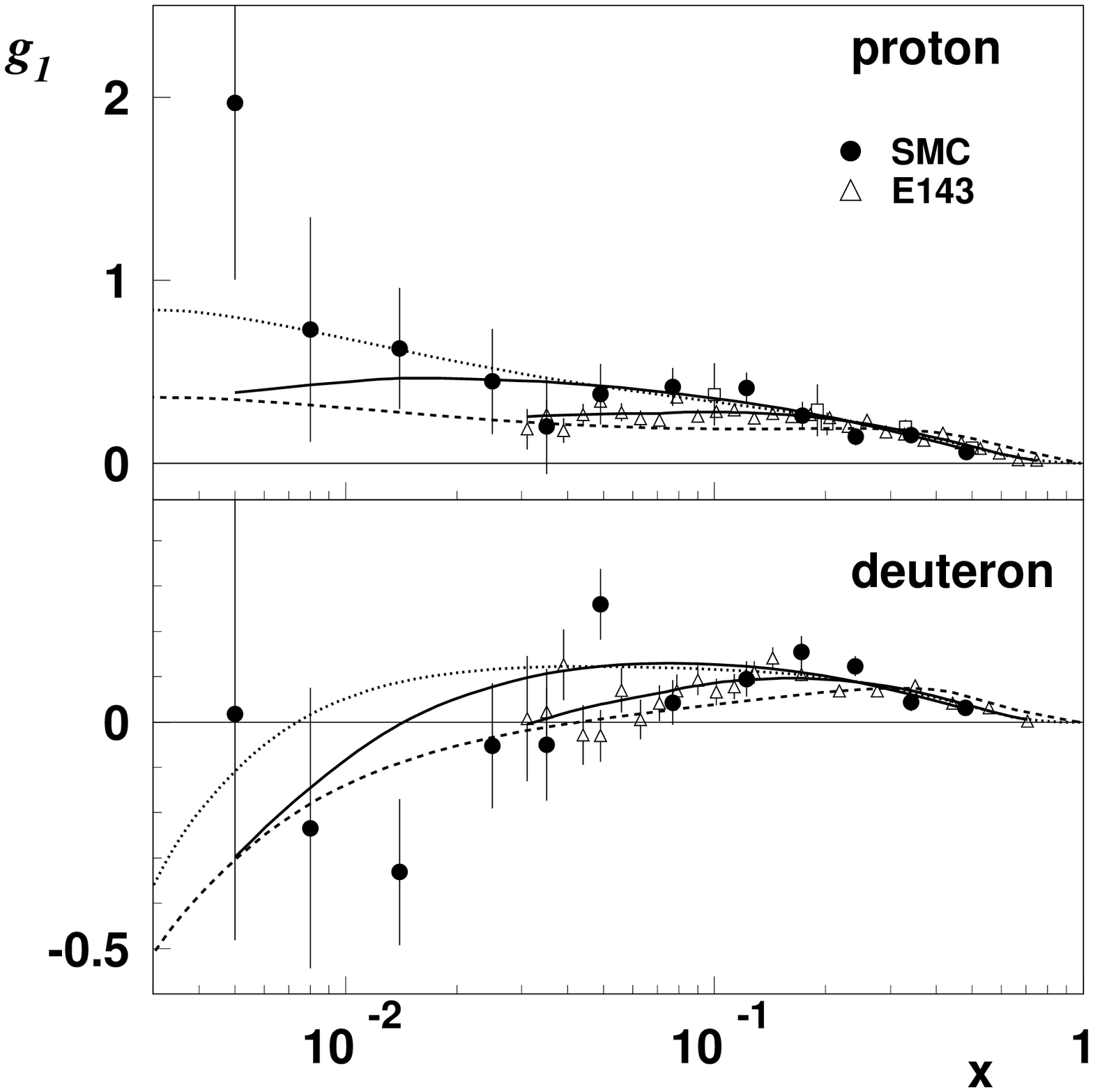}}
\caption{
NLO QCD fits\protect\cite{SMC96z} to the SMC and E-143 proton (top) and deuteron 
data (bottom) for the $Q^2$ of the measurement (solid line), $Q^2=1~\GeVs$ (dashed),
and $Q^2=10~\GeVs$ (dotted). }
\label{fig:NLO}
\end{minipage}
\hfill
\begin{minipage}[t]{0.47\hsize}
\mbox{\epsfxsize=\hsize\epsfbox[20 10 538 538]{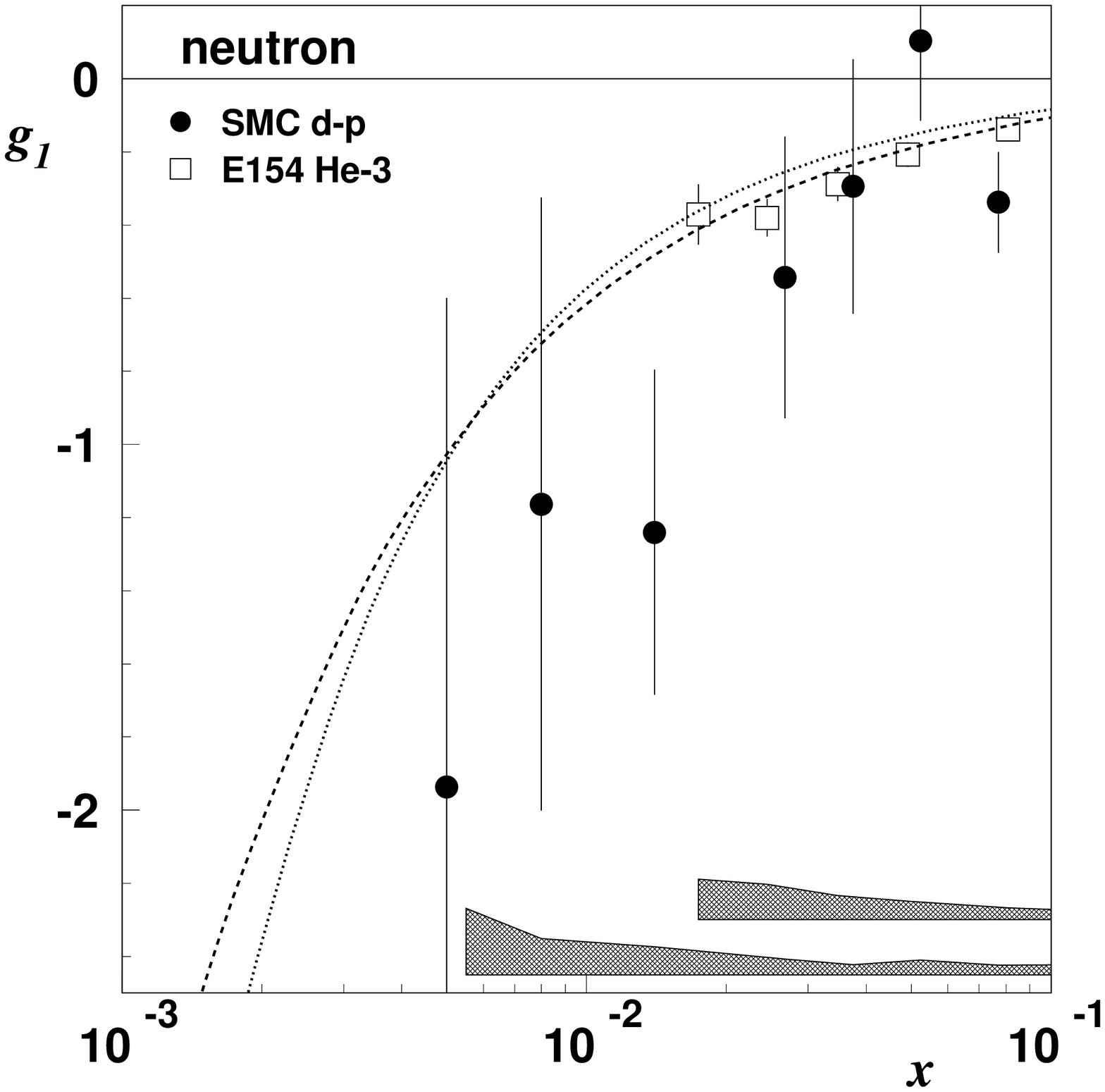}}
\caption{
The structure function $g_1$ of the neutron as a function of $x$ in the
small-$x$ region. Also shown are the QCD fits for $Q^2=1~\GeVs$ (dashed line)
and 10~GeV$^2$ (dotted line). The E-154 data are preliminary.}
\label{fig:g1n_sx}
\end{minipage}
\end{figure}

With more precise data available in a large kinematic range and after the
next-to-leading order
splitting functions were calculated~\cite{MeN95,Vog96a} in 1995 QCD analyses
of the $g_1$ data start to become sensible~\cite{GeS96a,GlR96,BaF95b,AlB96}.
From these analyses the size of the $Q^2$ dependence of $g_1$ can be estimated.
Using the method and code of Ref.~[\citen{BaF95b}] the SMC repeated\cite{SMC96z}
the QCD analysis using all EMC/SMC/E-143 proton and deuteron data 
(Fig.~\ref{fig:NLO}). Although the neutron data were not included in the fit,
the agreement is excellent (Fig.~\ref{fig:g1n_sx}).
A study of the systematic uncertainties showed that the main
error sources are the choice of the factorisation and renormalisation scales,
the parametrisations of the parton distribution functions, and the uncertainty
in the strong coupling constant. 
Due to these uncertainties the predicted $Q^2$ dependence in a given $x$ bin can vary
considerably, in particular in the small-$x$ region where 
the $Q^2$ slope of $g_1$ changes sign.
The systematic errors were accounted for in the evaluation of the sum rules by
the SMC.
A similar study including also the E-142 and
the preliminary E-154 neutron data was performed\cite{AlB96} by the authors
of the evolution code with similar results.

\begin{figure}[t]
\begin{minipage}[t]{0.49\hsize}
\begin{center}
\mbox{\epsfxsize=\hsize\epsfbox[30 10 538 546]{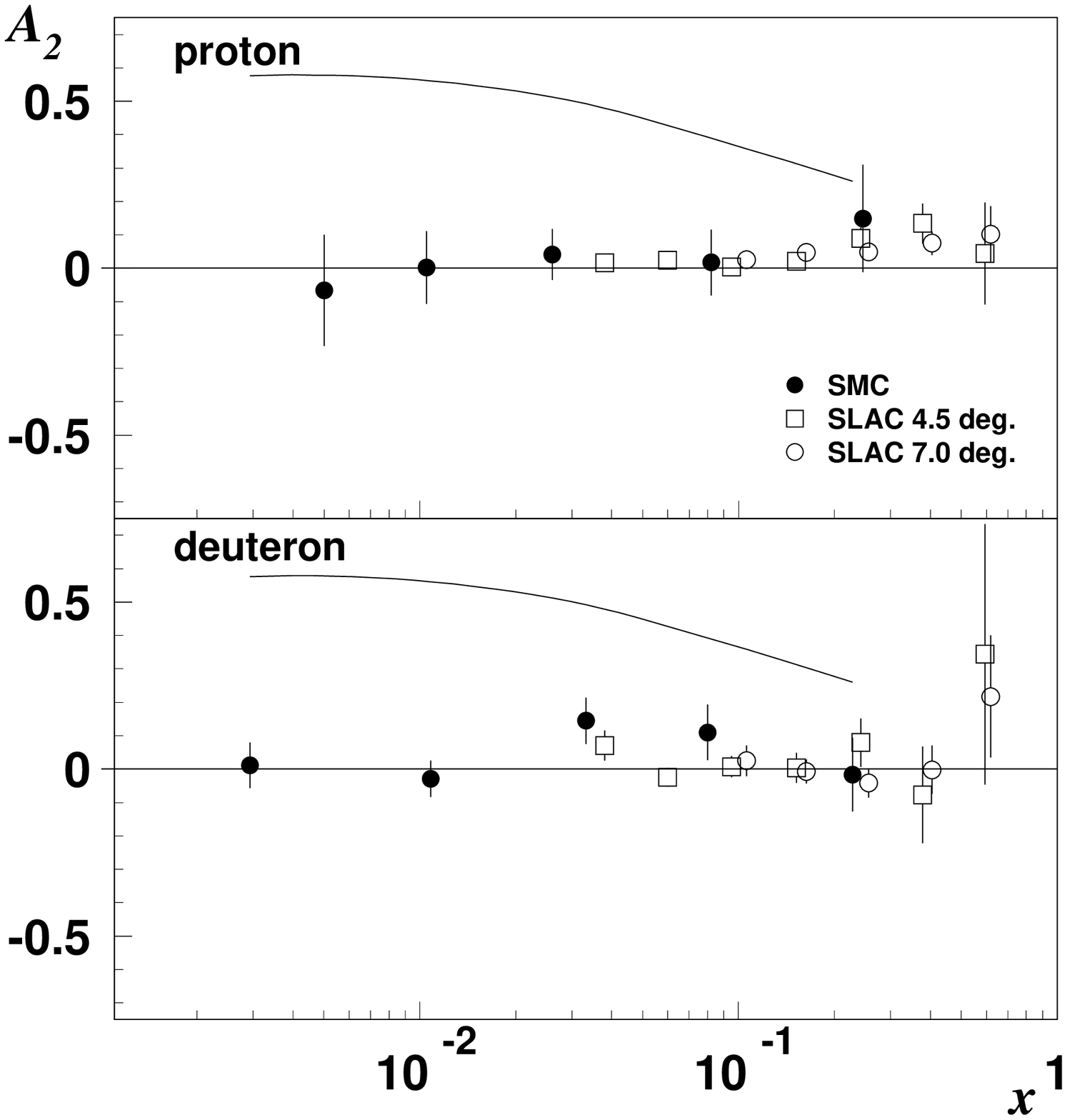}}
\end{center}

\caption{
The asymmetry $A_2$ as a function of $x$ for the proton and the deuteron
from the SMC and SLAC E-143 experiments.\protect\cite{SMC96z,SMC97a,E143_96a}.
Also shown is the limit $\protect\sqrt{R}$ for the kinematics of the SMC experiments
(solid line).}
\label{fig:a2}
\end{minipage}
\hfill
\begin{minipage}[t]{0.46\hsize}
\begin{center}
\mbox{\epsfxsize=\hsize\epsfbox[135 250 477 627]{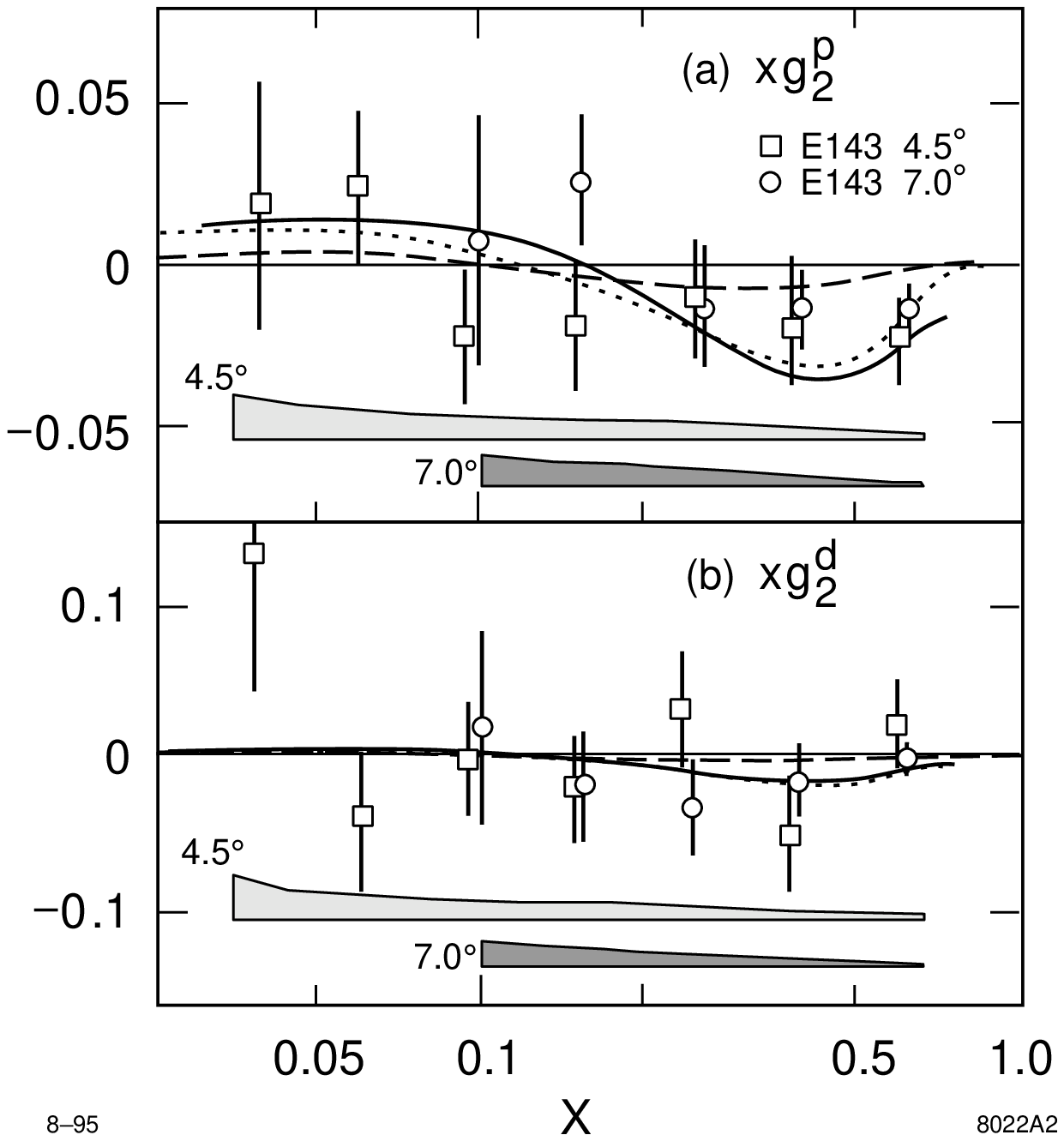}}
\end{center}

\caption{
The structure function $g_2$ as a function of $x$ for the proton and the deuteron
from the E-143 experiment.\protect\cite{E143_96a} The solid line indicates the
twist-2 part $g_2^{\rm ww}$. The dashed\protect\cite{SoM94} and 
dotted\protect\cite{Str93} lines refer to bag-model calculations of $g_2$.}
\label{fig:g2}
\end{minipage}
\end{figure}

The data for the asymmetry $A_2^{\rm p}$ and $A_2^{\rm d}$ from the 
SMC\cite{SMC94c,SMC96z,SMC97a} and the E-143 Collaboration\cite{E143_96a} are 
compatible with zero in the whole covered $x$ range with a
possible exception of $A_2^{\rm p}$ for $x>0.2$ (Fig.~\ref{fig:a2}).
The resulting structure functions, $g_2^{\rm p,d}$, are shown for the SLAC data in
Fig.~\ref{fig:g2}. The solid line in Fig.~\ref{fig:g2} represents the
twist-2 part\cite{WaW77}, $g_2^{\rm ww}$, which is calculable from $g_1$,
\beq
\label{eq:WW2}
g_2^{\rm ww}(x,Q^2) = -g_1(x,Q^2)+\int_x^1 \frac{g_1(y,Q^2)}{y}\,\id y.
\label{eq:WW_1}
\eeq
The data agree well with $g_2^{\rm ww}$ and a possible
twist-3 contribution, which relates to quark-gluon correlations\cite{Jaf90}, 
cannot be resolved within the accuracy of the present data.

\section{Test of the Sum Rules}
To evaluate the Ellis--Jaffe~\cite{ElJ74} and Bjorken~\cite{Bjo66} sum rules for 
the first moments of $g_1$,
\beq
\Gamma_1(Q^2_0)=\int_0^1 g_1(x,Q_0^2)\, \id x,
\label{eq:gamma1}
\eeq
the structure functions have to be known at a fixed, preferably high, value $Q_0^2$
of $Q^2$. 
Experiments cannot presently provide this kind of data because $Q^2$ is limited by the
incident lepton energy, $E$, to $Q^2\le 2MEx$. 
Thus at small $x$ the data necessarily are obtained at small $Q^2$. 
Until recently the standard procedure to evaluate $\Gamma_1$ from the asymmetries
was to assume scaling of either $A_1$ or of $g_1/F_1\simeq A_1$ and thus
to account only for the $Q^2$ evolution of $F_2$ and $R$ in Eq.~\ref{eq:g1}.

The second potentially dangerous step in the evaluation of $\Gamma_1$ is the 
extrapolation from the measured region
to $x=0$, i.e.\ from $x\ge0.03$ for the SLAC and $x\ge0.003$ for the SMC data.  
This extrapolation is usually performed assuming a Regge-type behaviour~\cite{Hei73} of the form 
$g_1\propto x^{-\alpha}$ with $-0.5<\alpha<0$\cite{ElK88}.
However, the theoretically proposed shapes vary widely~\cite{BaL94,ClR94} and the 
onset of a certain predicted behaviour is usually not well defined.
Therefore it is mandatory to measure to as small values of $x$ as possible.
In perturbative QCD one expects\cite{BaF95a} that $g_1$ turns negative at small $x$ and
reasonably high $Q^2$.
Such a drop is visible in the SMC neutron data and was recently confirmed 
by the preliminary E-154 neutron data\cite{E154_96a} for the region, 
$x>0.014$ (Fig.~\ref{fig:g1n_sx}).
This $x$ behaviour is  incompatible with the Regge-like behaviour of 
$g_1^{\rm n}$ for $x<0.03$ assumed in the analysis of the earlier E-142 data.
As is obvious from Fig.~\ref{fig:g1n_sx} it is difficult to perform a sensible
extrapolation to $x=0$ from the E-154 neutron data alone. The Collaboration only
quotes the integral for the measured region, $0.014<x<0.7$, of $-0.037\pm0.011$
at $Q^2=5~\GeVs$.
Assuming a power law the E-154 Collaboration found\cite{E154_96a} 
$g_1^{\rm n}(x\rightarrow0)=-0.02x^{-0.8}$ corresponding to an 
integral of $-0.043$ for  $0\le x<0.014$, which would even exceed the contribution 
from the measured region.

\begin{table}
\begin{center}
\caption{Results for the Ellis--Jaffe sum, $\Gamma_1(Q_0^2)$}
\label{tab:gamma1}

\begin{tabular}{@{}l@{}c|
                r@{\quad}r|
                r@{\quad}r|
                r@{\quad}r|
                r@{\quad}r@{}}
\hline
\hline
&\multicolumn{1}{l}{$\langle Q^2_0\rangle$\hskip1em~}&
\multicolumn{2}{c}{Proton} &
\multicolumn{2}{c}{Deuteron} &
\multicolumn{2}{c}{Neutron} &
\multicolumn{2}{c}{Bjorken} \\
\hline
\hline
 SMC         & 10 & 0.136 & (16) & 0.041 & (7) & $\it -0.047$ & \it (21) & 0.183 & (34) \\
 SMC$^{\rm scale}$& 10 & 0.139 & (17) & 0.037 & (8) & $\it -0.059$ & \it (24) & 0.198 & (35) \\
\hline
 E-143       &  3 & 0.127 & (11) & 0.042 & (5) & $\it -0.037$ & \it (14) & 0.163 & (19) \\
 E-142       &  2 &       &      &       &     & $    -0.031$ &     (11) &&\\
\hline
 \sc Hermes  &2.5 &       &      &       &     & $    -0.037$ &     (15) &&\\
\hline
 SMC$^{\rm comb}$&  5 & 0.142 & (11) & 0.038 & (6) & $    -0.061$ &     (16) & 0.202 & (22) \\
ABFR         &  3 & 0.114 &      & 0.027 &     & $    -0.056$ &          & 0.170 &      \\
\hline
Sum          & 10 & 0.171 & (4) & 0.071 & (4) & $    -0.017$ &      (4)  & 0.187 & (2) \\
\multicolumn{1}{c}{ rules}
             &  5 & 0.165 & (4) & 0.071 & (4) & $    -0.016$ &      (4) &  0.181 & (3) \\
             &  3 & 0.166 & (5) & 0.070 & (4) & $    -0.015$ &      (4) &  0.176 & (4) \\
\hline
\hline
\end{tabular}
\end{center}

\footnotesize 
The SMC and the ABFR~\protect\cite{AlB96} (Fit A) analyses use NLO QCD evolution. 
In the SMC$^{\rm scale}$ and the other analyses scaling of $A_1$ or $g_1/F_1$ is assumed.
The SMC$^{\rm comb}$ analysis includes SMC, E-143, and E-142 
data\protect\cite{SMC96z}.
The neutron values in {\it italics}\, are obtained by combining proton and 
deuteron data. Statistical and systematic errors were added in quadrature.
\end{table}

\begin{figure}[t]
  \begin{center}
      \mbox{\epsfxsize=0.65\hsize\epsfbox[40 0 394 538]{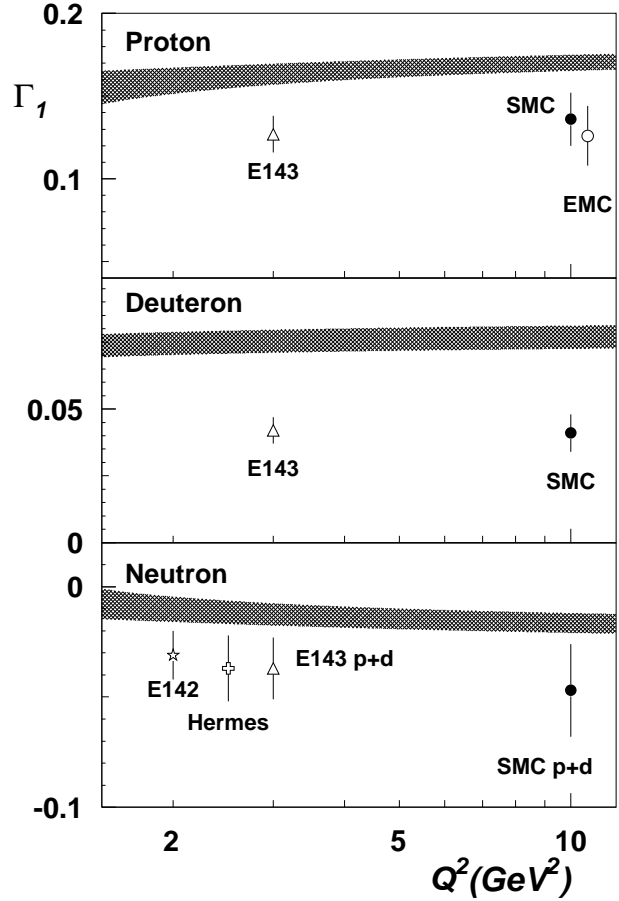}}

\caption{
      Results for the Ellis--Jaffe sum as a function of $Q^2$. The shaded
      bands indicate the theoretical prediction and its uncertainty. 
            }
   \label{fig:ejall}
   \end{center}
   \end{figure}
\begin{figure}[t]
  \begin{center}
      \mbox{\epsfxsize=0.8\hsize\epsfbox[69 35 589 483]{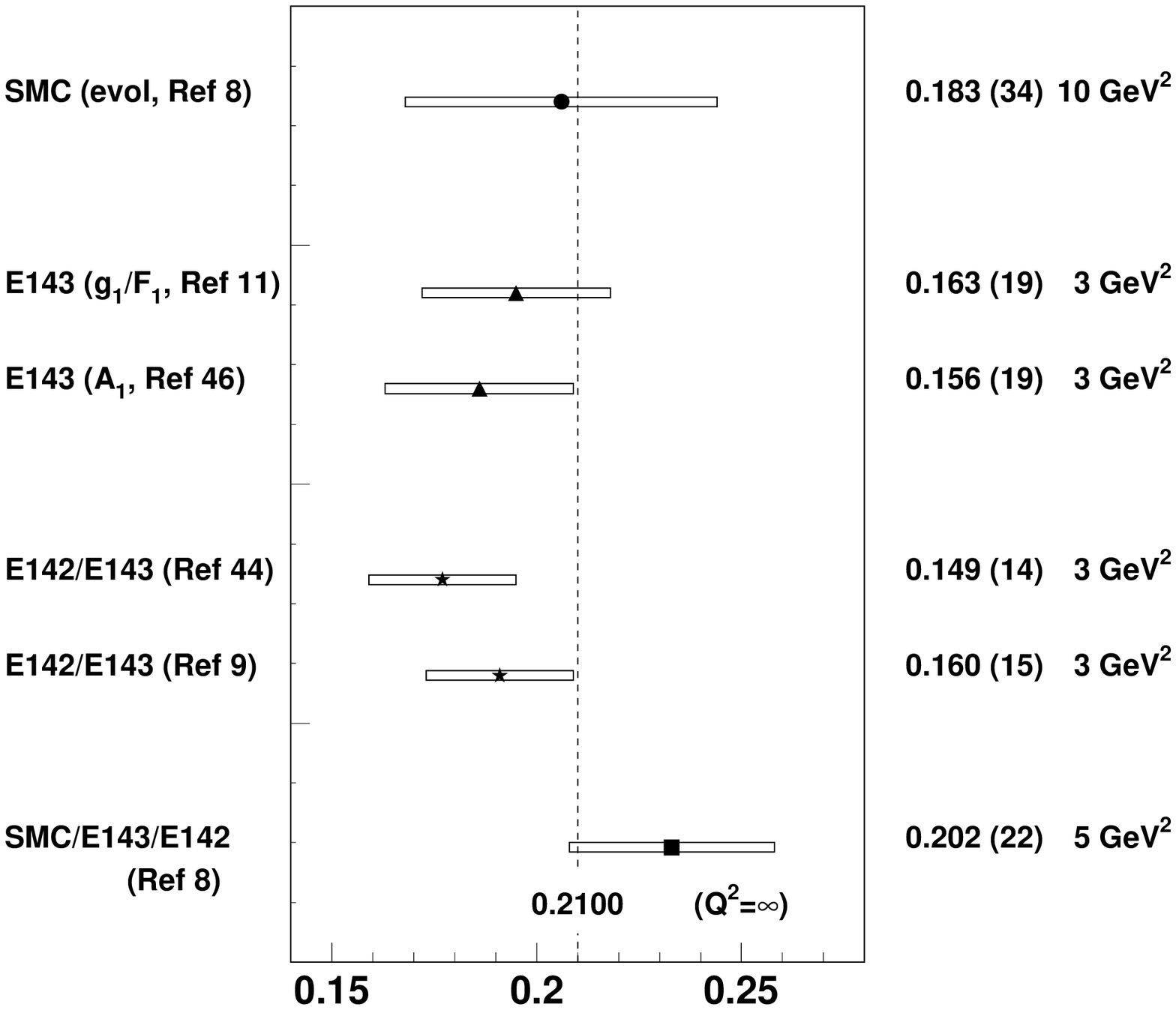}}

   \caption{
    Results for the Bjorken sum  for comparison evolved to $Q^2=\infty$.}
   \label{fig:bj2}
   \end{center}
   \end{figure}

The extrapolation to $x=1$ is uncritical due to the bound $|A_1|\le1$ and the 
smallness of $F_2$ in this region.
The results for the first moments of $g_1$ are summarised in 
Table~\ref{tab:gamma1} and are shown in Fig.~\ref{fig:ejall}.

The SMC was the first experimental group to include a full next-to-leading
order QCD evolution in their analyses\cite{SMC96z,SMC97a} (Figs.~\ref{fig:NLO},
\ref{fig:g1n_sx}).
The effect of the evolution from the $Q^2$ of the measurement, $Q_{\rm m}^2$,
to $Q^2_0$ was estimated by
\beq
g_1(x,Q^2_0) = g_1(x,Q^2_{\rm m}) +
               \left\{
               g_1^{\rm fit}(x,Q_0^2) - g_1^{\rm fit}(x,Q_{\rm m}^2)
               \right\},
\eeq
where $g_1^{\rm fit}$ refers to $g_1$ as calculated from the fitted
parton distribution functions.
The differences of the first moments obtained with QCD evolution and with
the scaling assumption for $A_1$ are small (Table~\ref{tab:gamma1}), partly 
due to cancellations of contributions from different $x$ regions.
The Ellis--Jaffe predictions given in Table~\ref{tab:gamma1}
and shown as shaded bands in Fig.~\ref{fig:ejall}
were calculated using QCD corrections~\cite{Lar94} up to order 
${\cal O}(\alpha_s^2)$ with three quark flavours,
$\alpha_s(M_{\rm Z}^2)=0.118\pm0.003$\cite{PDG96},
$g_a=1.2601\pm0.0025$\cite{PDG96}, and $F/D=0.575\pm0.016$\cite{ClR93}.
All experiments find a violation of the Ellis--Jaffe sum rule independent of 
the target material.
The most significant results are obtained for the deuteron, where the violation
amounts to 4.5 (3.7) standard deviations for the E-143 (SMC) data. 
Also the re-analysed E-142 neutron data~\cite{E142_96} now show a violation of
the sum rule, while when first published~\cite{E142_93} in 1993 agreement 
was reported.

The results for the Bjorken sum, $\Gamma_1^{\rm p}-\Gamma_1^{\rm n}$,
are summarised in Table~\ref{tab:gamma1} and Fig.~\ref{fig:bj2}, where for
comparison all experimental results were evolved to $Q^2=\infty$ using 
corrections~\cite{LaV91} up to order ${\cal O}(\alpha_s^3)$ and the constants 
given above.
All results are in agreement with the Bjorken sum rule prediction,
$0.2100\pm0.0003$, for $Q^2=\infty$. The value for the E-143 data assuming
scaling of $A_1$ instead of $g_1/F_1$ was taken from Ref.~\citen{McR96}.
For the SLAC data obtained at $Q^2=3~\GeVs$ the agreement improves when
higher-order corrections are estimated using the technique of Pad\'e
approximants~\cite{ElG96}.
For small $Q^2$ this method yields a much stronger dependence of the Bjorken 
sum rule on $\alpha_s$ than obtained from the corrections up to 
${\cal O}(\alpha_s^3)$. 
As a consequence the determination\cite{ElG96} of $\alpha_s$ 
from the Bjorken sum data at small $Q^2$ results in a very small statistical error,
$\alpha_s(M_{\rm Z}^2)=0.117^{+0.004}_{-0.007}\pm0.002$.
This method would yield similarly small errors when applied to e.g.\ the
CCFR data~\cite{CCFR95a} for the Gross--Llewelyn Smith sum.
The assessment of the theoretical uncertainties shown as second error for $\alpha_s$ 
has been criticised~\cite{For96z}.
Using corrections to ${\cal O}(\alpha_s^3)$ only, one obtains\cite{AlB96} about 
three-times larger uncertainties for the strong coupling constant.
In this case a more significant result can be obtained\cite{AlB96} from the scaling 
violations of $g_1$, $\alpha_s(M_{\rm Z}^2)=0.120^{+0.010}_{-0.008}$.

The SMC performed a combined analysis\cite{SMC96z} of the first moments
at $Q^2=5~\mbox{GeV}^2$ using a next-to-leading order QCD evolution. 
The analysis was performed $x$ bin per $x$ bin  and includes all published 
proton, deuteron, and neutron data.
In such a procedure the SLAC extrapolations for $0.003<x<0.03$ are replaced
by the SMC small-$x$ data. The result,
\beq
\Gamma^{\rm p}-\Gamma^{\rm n}=0.202\pm0.022\hskip1cm\mbox{at}\hskip0.3cmQ^2=5~\mbox{GeV}^2,
\eeq
lies somewhat above the theoretical prediction of $0.181\pm0.003$.
The combined result is larger than the individual results because
the SMC data yield a larger contribution at small $x$ than the SLAC extrapolations,
while at large $x$ the SLAC data lie slightly above the SMC data.
In better agreement with the Bjorken sum rule is the integral of the fitted
structure functions from the QCD analysis, which were used to evolve the data,
$\int (g_1^{\rm p}-g_1^{\rm n})^{\rm fit}\,\id x=0.188$ at $Q^2=5~\mbox{GeV}^2$. 
However, this result depends on the parametrisation of the parton distribution 
functions, particularly in the unmeasured region. 
In a similar fit\cite{AlB96} including also the neutron data
the authors of the evolution code found a Bjorken integral of  
0.170 at $Q^2=3~\GeVs$ (Fit A) also in good agreement with the Bjorken sum 
rule.
A significant difference between the two methods is that the SMC combined analysis
involves a Regge-type extrapolation, while in the integration method the fitted $g_1$
is also used in the region $x<0.003$. 

\begin{table}
\begin{center}
\caption{Results for $\Delta\Sigma$ and $\Delta s$ at $Q^2=\infty$}
\label{tab:DSigma}
\begin{tabular}{lcr@{\quad}rr@{\quad}r}
\hline
\hline
Exp. & target & \multicolumn{2}{c}{$\Delta\Sigma$} &\multicolumn{2}{c}{$\Delta s$}\\
\hline
SMC   & p & 0.25 &(15)  & $-0.11 $ & (5) \\
SMC   & d & 0.28 & (7)  & $-0.10 $ & (2) \\
E-143 & p & 0.22 &(10)  & $-0.12 $ & (3) \\
E-143 & d & 0.30 & (5)  & $-0.09 $ & (2) \\
E-142 & n & 0.41 &(13)  & $-0.06 $ & (4) \\
\hline
SMC$^{\rm comb}$ 
      &   & 0.28 &(6)   & $-0.11 $ & (2) \\
\hline
ABFR  &   & 0.10 & (7)  &          &     \\
\hline
\hline
\end{tabular}
\end{center}
\end{table}

The results for the flavour-singlet axial-current matrix element \footnote{Note that 
the notation $\Delta\Sigma$ corresponds to $a_0(\infty)$ in e.g.\ 
Refs.~[\protect\citen{AlB96,SMC96z}].}, 
$\Delta\Sigma$, 
and the polarisation of the strange quarks, $\Delta s=\int \Delta s(x)\,\id x$,
are summarised in Table~\ref{tab:DSigma}. 
The values were obtained from the results for $\Gamma_1$ using the constants 
given above. 
The dependence of $\Delta\Sigma$ on the SU(3) flavour symmetry assumed in the
derivation of the Ellis--Jaffe sum rule is weak\cite{LiL95}, 
while the result for $\Delta s$ strongly depends on it.
All results are given for $Q^2=\infty$. The SMC combined result,
\beq
\Delta\Sigma=0.28\pm0.06 \hskip 1cm \mbox{and} \hskip 1cm \Delta s=-0.11\pm0.002,
\eeq
was evaluated in the same way as the combined value for the Bjorken sum and includes
the SMC, E-142, and E-143 data.
For the polarisation of the up and down quarks the SMC combined analysis yields
$\Delta u=0.81\pm0.02$ and $\Delta d=-0.42\pm0.02$.
A much smaller value for $\Delta\Sigma$ is obtained from QCD fits to the data, 
due to the different treatment of the unmeasured small $x$ region.
For the most significant case, the deuteron, the extrapolation from $x=0.003$ to
$x=0$ amounts to $-0.016$ for the fit (ABFR\cite{AlB96} in Table~\ref{tab:DSigma}) 
at $Q^2_0=10~\GeVs$, while a Regge-like extrapolation from the SMC deuteron data 
yields zero. The question of the small-$x$ behaviour of $g_1$ must eventually be
settled by a measurement.

\begin{figure}[t]
\begin{minipage}[t]{0.59\hsize}
\begin{center}
\mbox{\epsfxsize=\hsize\epsfbox[34 165 503 643]{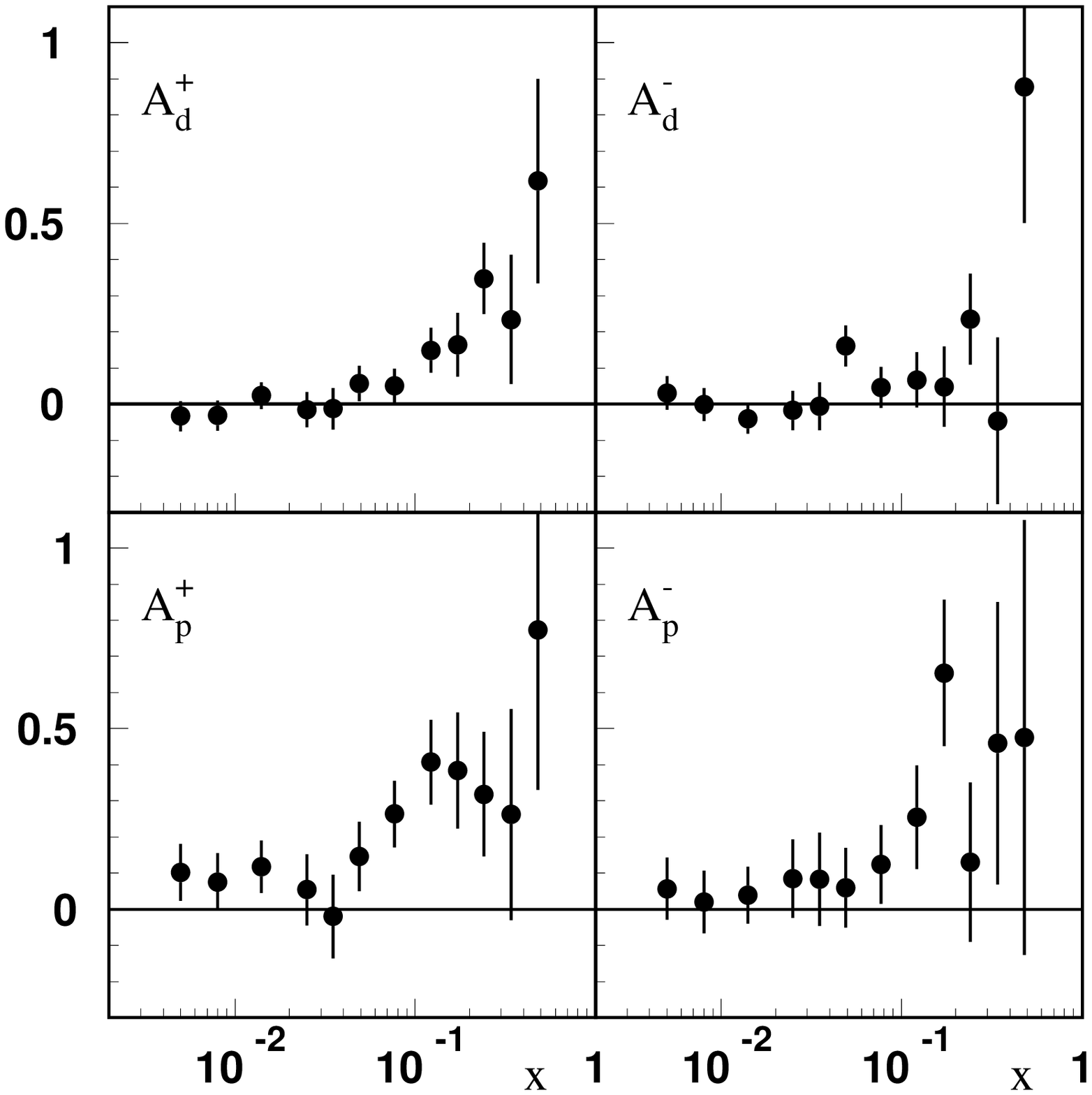}}
\end{center}

\caption{The asymmetries for positive (left) and negative(right) hadrons
for deuteron (top) and proton (bottom) targets as a function of $x$.
}
\label{fig:hadasy}
\end{minipage}
\hfill
\begin{minipage}[t]{0.37\hsize}
\begin{center}
\mbox{\epsfxsize=\hsize\epsfbox[90 28 406 539]{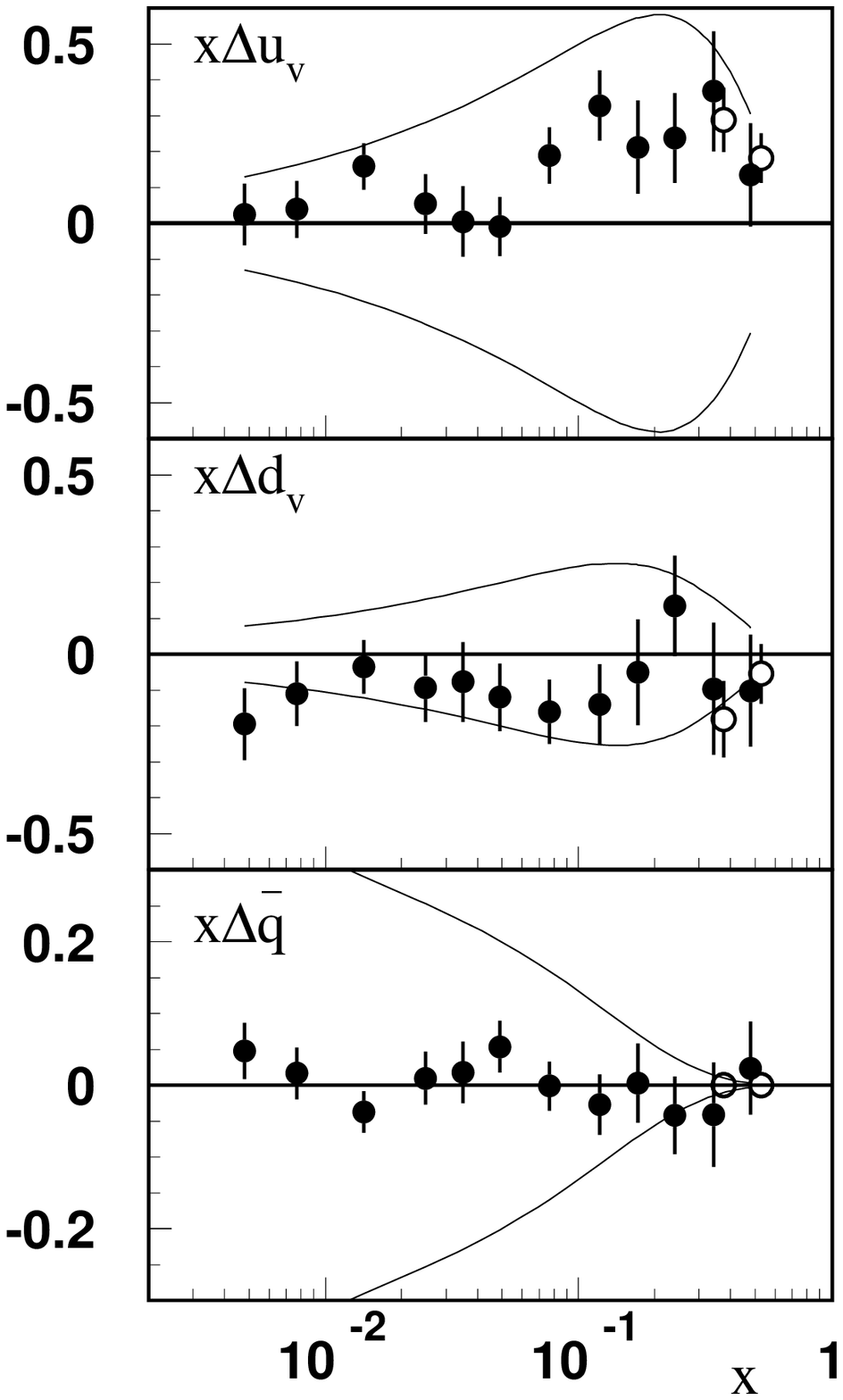}}
\end{center}

\caption{
Preliminary SMC data for the
polarisation of the valence up and down quarks and the light antiquarks.
}
\label{fig:Dqval}
\end{minipage}
\end{figure}

\section{Semi-inclusive Asymmetries}

In unpolarised deep inelastic scattering the flavour separation of the parton 
distribution functions is accomplished by the combination of data from charged-lepton
and charged-current neutrino scattering, which weight the individual quark
flavours differently.
Neutrino experiments require huge targets, which are not available in the polarised 
case. If the HERA proton beam can be polarised, charged-current electron-proton
scattering at high $Q^2$  would become possible.\cite{AnG97a}
Presently the only access to flavour separation is semi-inclusive scattering, 
where the emerging hadron carries some memory of the flavour of the initially 
hit quark. The fragmentation of a quark $q$ into a hadron $h$ is described by the 
fragmentation functions $D_q^h$. The fragmentation factorises from the hard scattering
process and the differential cross section reads,
\beq
\frac{1}{\sigma^\mu}\frac{\id\sigma^\pm}{\id z}
=
\frac{1}{N^\mu}\frac{\id N^\pm}{\id z}
=
\frac{\sum_q e_q^2 \, q(x,Q^2) \, D_q^{h^\pm}(z,Q^2)}
     {\sum_q e_q^2\, q(x,Q^2)},
\eeq
where $N^\pm$ is the number of produced hadrons,
$N^\mu$ the is the number of deep-inelastic events,
$z=E_h/\nu$ is the quark's energy fraction carried by the hadron,
and $\sigma^\pm$ and $\sigma^\mu$ are the inclusive and semi-inclusive
differential cross sections.
The sums run over $q= \rm u,d,s,\bar u,\bar d,\bar s$ and
$e_q$ are the quark charges.

The only data on flavour separation~\cite{SMC96a} to date come from the SMC.
The preliminary data presented here include the 1995 data set.
The asymmetries for positive and negative hadrons for both, proton and deuteron
targets, are shown is Fig.~\ref{fig:hadasy}.
In the analysis a cut of $z>0.2$ was applied and the fragmentation functions were
taken from EMC measurements.\cite{EMC89b}
The HERMES Collaboration also presented\cite{HER96b} hadron asymmetries for a 
$^3$He target. However, only together with their not yet available 1996 proton data
a flavour separation can be performed.
The preliminary SMC results for the valence quark distributions, $\Delta q_v$ with
$q_v(x)=q(x)-\overline{q}(x)$, are shown in Fig.~\ref{fig:Dqval}.
The up valence quarks show a positive polarisation over the whole $x$ range, while the
down valence quarks are polarised oppositely to the proton spin. 
For the first moments the SMC obtains
$
\Delta u_v = 0.85 \pm 0.14 \pm 0.12$,
$\Delta d_v = -0.58 \pm 0.16 \pm 0.11$,
and
$\Delta \overline{q} = 0.02 \pm 0.06 \pm 0.03$,
where $\Delta\overline{u}(x)=\Delta\overline{d}(x):=\Delta\overline{q}(x)$ was assumed.
The analysis is largely insensitive to $\Delta s(x)$, which was assumed to be 
proportional to $s(x)$ with a first moment as obtained from $\Gamma_1$.

\section{Conclusions and Outlook}

The main goals of the experiments initiated  after the discovery of the 
``EMC spin effect'',
in particular the confirmation of the EMC proton result, the extension of the
measurements to smaller $x$, a measurement of the neutron or deuteron structure
function and thus a test of the Bjorken sum rule are now largely achieved. 
The data from all experiments are in good agreement and draw a consistent 
picture, where the Bjorken sum rule is tested with a precision of about 10~\%
and where $\Delta\Sigma$ is of the order of 30~\%.
However, still the original question raised by the EMC result of how the 
spin of the nucleon
is built up from the spins and the orbital angular momenta of quarks and 
gluons is unanswered.
The interest has now turned to the $Q^2$ dependence of 
$g_1$ and the r\^ole of the gluon polarisation, $\Delta g$,
which via the axial anomaly~\cite{EfT88,AlR88,CaC88} contributes to 
$\Delta\Sigma$.
If the violation of the Ellis--Jaffe sum rules is entirely attributed to the
anomalous contribution of $\Delta g$ to $\Delta\Sigma$, a gluon polarisation
of $\Delta g=2.5$ at  $Q^2=10~\GeVs$ is required. QCD analyses indicate
that $\Delta g$ is positive and of about this size~\cite{AlB96,GlR96}.
Also a QCD sum rule calculation\cite{MaP96} indicates a large and positive
value for $\Delta g$.

To make further progress in our understanding of the nucleon's spin structure
a direct measurement of $\Delta g$ is indispensable. An unambiguous determination
of $\Delta g$ can only be achieved studying  a process where $\Delta g$ enters in
leading order. The cleanest such process is the photon-gluon fusion,
which can either be tagged by open charm, $\gamma^\star\rm g\rightarrow c\overline{c}$,
or 2+1-jet production\cite{H1_95a}.
The former is proposed for the approved COMPASS experiment at 
CERN~\cite{COMPASS96,Mal96a} while the later is aimed at by the project to polarise 
the HERA proton beam~\cite{heraws96}. Processes using hadronic probes,
like $\rm p\overline{p}$ collisions at RHIC~\cite{RSC92} are in general more
difficult to interpret. A detailed understanding of the nucleon's spin structure
must also include the third twist-2 structure function, 
the transversity\cite{JaJ92} $h_1$, which will be studied at HERMES, COMPASS,
and RHIC.

\section*{Acknowledgements}
I like to thank H.~Castilla for the inspiring atmosphere in Mexico City,
C.~W. de Jager, T.~J. Ketel, P.~J. Mulders, and J.~Oberski for
their hospitality in Amsterdam, and my colleagues in the SMC for the many 
discussions which were essential in  preparing these talks.

\begingroup
\renewcommand{\refout}[5]{%
   {\def\thearg{#1}\ifx\empty\thearg{}\else{\jf #1}\fi}%
   {\def\thearg{#2}\ifx\empty\thearg{}\else{~\bf#2}\fi}%
   {\def\thearg{#4}\ifx\empty\thearg{}\else{, #4}\fi}%
   {\def\thearg{#3}\ifx\empty\thearg{}\else{ (#3)}\fi}}%
\def\jf{\it}

\endgroup

\end{document}